\documentclass[11pt]{article}
\usepackage{axodraw}
\usepackage{epsfig}
\usepackage{amsfonts}
\usepackage{amsmath}
\usepackage{bbm}
\allowdisplaybreaks[1]

\input pix.sty

\renewcommand{\eq}{eq.~}
\renewcommand{\eqs}{eqs.~}
\renewcommand{\se}{sec.~}

\renewcommand{\fig}{fig.~}

\newcommand{\Nf}{N_{\rm f}}
\newcommand{\Nc}{N_{\rm c}}

\newcommand{\rmO}{{\mathcal{O}}}

\def\lsi{\raise0.3ex\hbox{$<$\kern-0.75em\raise-1.1ex\hbox{$\sim$}}}
\def\gsi{\raise0.3ex\hbox{$>$\kern-0.75em\raise-1.1ex\hbox{$\sim$}}}

\newcommand{\gsim}{\mathop{\gsi}}

\newcommand{\nF}{n_\rmii{F}}
\newcommand{\nB}{n_\rmii{B}}
 \renewcommand{\nF}[1]{{f}_\rmii{F{#1}}}
 \renewcommand{\nB}[1]{{f}_\rmii{B{#1}}}

\newcommand{\rmii}[1]{{\mbox{\tiny\rm{#1}}}}

\newcommand{\im}{\mathop{\mbox{Im}}}

\newcommand{\Tint}[1]{{\hbox{$\sum$}\!\!\!\!\!\!\!\int\,}_{\!\!\!\!\raise-0.9ex\hbox{$\scriptstyle{#1}$}}}
\newcommand{\Tinti}[1]{{{\Sigma}\!\!\!\!\raise0.3ex\hbox{$\int$}_\rmii{${#1}$}}}

\newcommand{\unit}{{\mathbbm{1}}} 
\newcommand{\bi}{\begin{itemize}}
\newcommand{\ei}{\end{itemize}}

\newcommand{\hide}[1]{ }
\newcommand{\bsl}[1]{\,\slash\!\!\!\!{#1}\,}

\newcommand{\ff}{\rmi{\sl f\,}}
\newcommand{\nsum}{{\textstyle{\sum}\,}}
\newcommand{\deltabar}{\raise-0.02em\hbox{$\bar{}$}\hspace*{-0.8mm}{\delta}}
\def\TAsc(#1,#2)(#3,#4,#5)%
{\SetWidth{2.0}\CArc(#1,#2)(#3,#4,#5)\SetWidth{1.0}}
\def\Lwidth{3}

\def\TAgl(#1,#2)(#3,#4,#5){\SetWidth{2.0}\PhotonArc(#1,#2)(#3,#4,#5){\Lwidth}%
{6.283 #3 mul 360 div #4 #5 sub #4 #5 sub mul sqrt mul Tdensity mul}%
\SetWidth{1.0}}
\def\TLgl(#1,#2)(#3,#4){\SetWidth{2.0}\Photon(#1,#2)(#3,#4){\Lwidth}
{#1 #3 sub #1 #3 sub mul #2 #4 sub #2 #4 sub mul add sqrt Tdensity mul}%
\SetWidth{1.0}}

\renewcommand{\picc}[1]{\;\parbox[c]{60pt}{\begin{picture}(60,30)(0,0)
\SetWidth{1.0}\SetScale{1.3} #1 \end{picture}}\; }
\def\Lwidth{1.3}

\newcommand{\picu}[1]{\;\parbox[c]{30pt}{\begin{picture}(30,30)(0,0)
\SetWidth{1.0}\SetScale{1.0} #1 \end{picture}}\; }
\def\EleA{\picu{%
 \CArc(15,15)(15,0,360)%
 \CArc(15,15)(17,0,360)%
 \Lgl(15,0)(15,30)%
 \COval(15,-1)(2,2)(0){Black}{Black}%
 \COval(15,31)(2,2)(0){Black}{Black}%
}}
\def\EleB{\picu{%
 \CArc(15,15)(15,0,360)%
 \CArc(15,15)(17,0,360)%
 \Lgl(15,0)(15,30)%
 \COval(15,-1)(2,2)(0){Black}{Black}%
 \COval(15,31)(2,2)(0){Black}{Black}%
 \Agl(29,15)(8,100,260)%
}}
\def\EleBB{\picu{%
 \CArc(15,15)(15,0,360)%
 \CArc(15,15)(17,0,360)%
 \Lgl(15,0)(15,30)%
 \Lgl(0,15)(12,15)%
 \Lgl(18,15)(30,15)%
 \COval(15,-1)(2,2)(0){Black}{Black}%
 \COval(15,31)(2,2)(0){Black}{Black}%
}}
\def\EleC{\picu{%
 \CArc(15,15)(15,0,360)%
 \CArc(15,15)(17,0,360)%
 \Agl(7,4)(8,-30,130)%
 \Agl(23,26)(8,150,310)%
 \COval(15,-1)(2,2)(0){Black}{Black}%
 \COval(15,31)(2,2)(0){Black}{Black}%
}}
\def\EleD{\picu{%
 \CArc(15,15)(15,0,360)%
 \CArc(15,15)(17,0,360)%
 \Agl(23,4)(8,50,210)%
 \Agl(23,26)(8,150,310)%
 \COval(15,-1)(2,2)(0){Black}{Black}%
 \COval(15,31)(2,2)(0){Black}{Black}%
}}
\def\EleE{\picu{%
 \CArc(15,15)(15,0,360)%
 \CArc(15,15)(17,0,360)%
 \Agl(40,30)(25,180,242)%
 \Agl(40,0)(25,118,137)%
 \Agl(40,0)(25,150,180)%
 \COval(15,-1)(2,2)(0){Black}{Black}%
 \COval(15,31)(2,2)(0){Black}{Black}%
}}
\def\EleF{\picu{%
 \CArc(15,15)(15,0,360)%
 \CArc(15,15)(17,0,360)%
 \Lgl(15,0)(15,30)%
 \Agl(40,40)(26,205,245)%
 \COval(15,-1)(2,2)(0){Black}{Black}%
 \COval(15,31)(2,2)(0){Black}{Black}%
}}
\def\EleG{\picu{%
 \CArc(15,15)(15,0,360)%
 \CArc(15,15)(17,0,360)%
 \Agl(40,15)(30,150,210)%
 \Agl(-10,15)(30,-30,30)%
 \COval(15,-1)(2,2)(0){Black}{Black}%
 \COval(15,31)(2,2)(0){Black}{Black}%
}}
\def\EleH{\picu{%
 \CArc(15,15)(15,0,360)%
 \CArc(15,15)(17,0,360)%
 \Lgl(15,0)(15,12)%
 \Agl(10,21.5)(10,-65,65)%
 \Agl(20,21.5)(10,115,245)%
 \COval(15,-1)(2,2)(0){Black}{Black}%
 \COval(15,31)(2,2)(0){Black}{Black}%
}}
\def\EleI{\picu{%
 \CArc(15,15)(15,0,360)%
 \CArc(15,15)(17,0,360)%
 \Lgl(15,0)(15,30)%
 \Lgl(15,15)(30,15)%
 \COval(15,-1)(2,2)(0){Black}{Black}%
 \COval(15,31)(2,2)(0){Black}{Black}%
}}
\def\EleJ{\picu{%
 \CArc(15,15)(15,0,360)%
 \CArc(15,15)(17,0,360)%
 \Lgl(15,0)(15,30)%
 \COval(15,-1)(2,2)(0){Black}{Black}%
 \COval(15,31)(2,2)(0){Black}{Black}%
 \GCirc(15,15){4}{0.5}
}}
\def\EleK{\picu{%
 \CArc(15,27)(5,0,360)%
 \CArc(15,3)(5,0,360)%
 \CArc(15,27)(7,0,360)%
 \CArc(15,3)(7,0,360)%
 \Agl(20,15)(10,142,218)%
 \Agl(10,15)(10,-38,38)%
 \COval(12,8)(2,2)(0){Black}{Black}%
 \COval(12,22)(2,2)(0){Black}{Black}%
}}
\def\EleL{\picu{%
 \CArc(15,27)(5,0,360)%
 \CArc(15,3)(5,0,360)%
 \CArc(15,27)(7,0,360)%
 \CArc(15,3)(7,0,360)%
 \Agl(20,15)(10,142,218)%
 \Agl(10,15)(10,-38,38)%
 \COval(12,8)(2,2)(0){Black}{Black}%
 \COval(18,22)(2,2)(0){Black}{Black}%
}}
\newcommand{\piB}[1]{\;\parbox[c]{60pt}{\begin{picture}(60,40)(0,0)
\SetWidth{1.0}\SetScale{1.0} #1 \end{picture}}\;}
\def\ScatA{\piB{%
 \SetWidth{2.0} 
 \Lqu(10,30)(11,29)%
 \Laqu(10,10)(11,11)%
 \SetWidth{1.0} 
 \Line(-1.5,40)(18.5,20)%
 \Line(0,41.5)(21.5,20)%
 \Line(0,-1.5)(21.5,20)%
 \Line(-1.5,0)(18.5,20)%
 \Lgl(21.5,20)(40,20)%
 \Lgl(40,20)(60,40)%
 \Lgl(40,20)(60,0)%
}}
\def\ScatB{\piB{%
 \SetWidth{2.0} 
 \Lqu(15,35)(16,34)%
 \Laqu(15,5)(16,6)%
 \SetWidth{1.0} 
 \Line(8.5,40)(21.5,27)%
 \Line(10,41.5)(23.5,28)%
 \Line(10,-1.5)(23.5,12)%
 \Line(8.5,0)(21.5,13)%
 \Line(21.5,13)(21.5,27)%
 \Line(23.5,12)(23.5,28)%
 \Lgl(23.5,28)(37,41.5)%
 \Lgl(23.5,12)(37,-1.5)%
}}
\def\ScatC{\piB{%
 \SetWidth{2.0} 
 \Lqu(15,35)(16,34)%
 \Laqu(15,5)(16,6)%
 \SetWidth{1.0} 
 \Line(8.5,40)(21.5,27)%
 \Line(10,41.5)(23.5,28)%
 \Line(10,-1.5)(23.5,12)%
 \Line(8.5,0)(21.5,13)%
 \Line(21.5,13)(21.5,27)%
 \Line(23.5,12)(23.5,28)%
 \Lgl(23.5,12)(29.5,18)%
 \Lgl(33.5,22)(53,41.5)%
 \Lgl(23.5,28)(53,-1.5)%
}}
\def\ScatD{\piB{%
 \SetWidth{2.0} 
 \Lqu(10,30)(11,29)%
 \Laqu(10,10)(11,11)%
 \SetWidth{1.0} 
 \Line(-1.5,40)(18.5,20)%
 \Line(0,41.5)(21.5,20)%
 \Line(0,-1.5)(21.5,20)%
 \Line(-1.5,0)(18.5,20)%
 \Lgl(21.5,20)(40,20)%
 \Laqu(40,20)(60,40)%
 \Lqu(40,20)(60,0)%
}}
\def\ScatDe{\piB{%
 \SetWidth{2.0} 
 \Lqu(30,40.5)(31,40.5)%
 \Laqu(30,-0.5)(31,-0.5)%
 \SetWidth{1.0} 
 \CArc(30,12)(30,18,162)%
 \CArc(30,12)(28,18,162)%
 \CArc(30,28)(30,198,342)%
 \CArc(30,28)(28,198,342)%
 \Lgl(5,20)(20,20)%
 \Lgl(40,20)(55,20)%
 \Aqu(30,20)(10,0,180)%
 \Aqu(30,20)(10,180,360)%
 \COval(2.5,20)(2,2)(0){Black}{Black}%
 \COval(57.5,20)(2,2)(0){Black}{Black}%
 \SetWidth{0.5}
 \LongArrow(-25,20)(-1,20)%
 \LongArrow(61,20)(85,20)%
 \Text(-8,28)[r]{\bf $(\omega_n,\vec{0})$}%
 \Text(68,28)[l]{\bf $(\omega_n,\vec{0})$}%
 \Line(15,-10)(45,50)
}}
\def\ScatDl{\piB{%
 \SetWidth{2.0} 
 \Lqu(5,35)(6,34)%
 \Laqu(5,5)(6,6)%
 \SetWidth{1.0} 
 \Line(-1.5,40)(18.5,20)%
 \Line(0,41.5)(21.5,20)%
 \Line(0,-1.5)(21.5,20)%
 \Line(-1.5,0)(18.5,20)%
 \Lgl(21.5,20)(40,20)%
 \Agl(20,20)(10,139,221)%
 \Lqu(40,20)(60,40)%
 \Laqu(40,20)(60,0)%
}}
\def\ScatDk{\piB{%
 \SetWidth{2.0} 
 \Lqu(5,35)(6,34)%
 \Laqu(5,5)(6,6)%
 \SetWidth{1.0} 
 \Line(-1.5,40)(18.5,20)%
 \Line(0,41.5)(21.5,20)%
 \Line(0,-1.5)(21.5,20)%
 \Line(-1.5,0)(18.5,20)%
 \Lgl(21.5,20)(40,20)%
 \Agl(30,10)(20,45,131)%
 \Line(40,20)(45,25)%
 \Lqu(45,25)(60,40)%
 \Line(40,20)(45,15)%
 \Laqu(45,15)(60,0)%
}}
\def\ScatDm{\piB{%
 \SetWidth{2.0} 
 \Lqu(10,30)(11,29)%
 \Laqu(10,10)(11,11)%
 \SetWidth{1.0} 
 \Line(-1.5,40)(18.5,20)%
 \Line(0,41.5)(21.5,20)%
 \Line(0,-1.5)(21.5,20)%
 \Line(-1.5,0)(18.5,20)%
 \Lgl(21.5,20)(40,20)%
 \Agl(40,20)(10,-45,45)%
 \Line(40,20)(45,25)%
 \Lqu(45,25)(60,40)%
 \Line(40,20)(45,15)%
 \Laqu(45,15)(60,0)%
}}

\makeatletter \@addtoreset{equation}{section} \makeatother

\makeatletter
\renewcommand\section{\@startsection {section}{1}{\z@}%
                                   {-5.5ex \@plus -1ex \@minus -.2ex}
                                   {2.3ex \@plus.2ex}%
                                   {\normalfont\large\bfseries}}
\renewcommand\subsection{\@startsection{subsection}{2}{\z@}%
                                     {-3.25ex\@plus -1ex \@minus -.2ex}%
                                     {1.5ex \@plus .2ex}%
                                     {\normalfont\normalsize\bfseries}}
\renewcommand\thesection {\@arabic\c@section}
\renewcommand\thesubsection   {\thesection.\@arabic\c@subsection}
\renewcommand{\@seccntformat}[1]{%
\csname the#1\endcsname.\hspace{1.0em}}
\makeatother

\begin{document}

\flushbottom

\begin{titlepage}

\begin{flushright}
BI-TP 2012/15\\
INT-PUB-12-026\\ 
\vspace*{1cm}
\end{flushright}
\begin{centering}
\vfill

{\Large{\bf
 Heavy quark chemical equilibration rate \\[2mm]
 as a transport coefficient
}} 

\vspace{0.8cm}

D.~B\"odeker$^{\rm a}$,  
M.~Laine$^{\rm b}$ 

\vspace{0.8cm}

$^\rmi{a}$%
{\em
Faculty of Physics, University of Bielefeld, 
D-33501 Bielefeld, Germany\\}

\vspace*{0.3cm}

$^\rmi{b}$%
{\em
Institute for Theoretical Physics, 
Albert Einstein Center, University of Bern, \\ 
Sidlerstrasse 5, CH-3012 Bern, Switzerland\\}

\vspace*{0.8cm}

\mbox{\bf Abstract}
 
\end{centering}

\vspace*{0.3cm}
 
\noindent
Motivated by indications that heavy (charm and bottom)
quarks interact strongly at temperatures generated in heavy
ion collision experiments, we suggest a non-perturbative
definition of a heavy quark chemical equilibration rate as
a transport coefficient. Within leading-order perturbation
theory (corresponding to 3-loop level), the definition is 
argued to reduce to an expression obtained from the Boltzmann 
equation. Around $T \sim 400$~MeV, an order-of-magnitude estimate 
for charm yields a rate $\Gamma^{-1}_\rmi{chem} \gsim 60$~fm/c 
which remains too slow to play a practical role in current 
experiments. However, the rate increases rapidly with $T$ 
and, due to non-linear effects, also if the initial state 
contains an overabundance of heavy quarks.

\vfill

 
\vspace*{1cm}
  
\noindent
July
 2012

\vfill

\end{titlepage}

%
\section{Introduction}

In a fully thermalized medium, the momenta of bosons and fermions 
are distributed according to the Bose and Fermi distributions, 
respectively, parametrized by a single temperature, $T$, and chemical 
potentials associated with conserved global charges. In contrast, 
the most important cosmological relics, such as Light Element
Abundances, Dark Matter, or 
Baryon Asymmetry, rely on deviations from thermal equilibrium. 
In a canonical Dark Matter scenario, for instance, 
the overall abundance of the Dark Matter particles is determined
through a ``freeze-out'' period, which takes place when their annihilation 
rate becomes too slow to track the total number density 
determined by the Fermi distribution, which decreases exponentially
when $\pi T \ll M$, where $M$ denotes the particle mass.  
Since the number densities of particles
and antiparticles remain equal, this deviation cannot 
in relativistic field theory be represented
through a chemical potential, and we speak of chemical non-equilibrium. 
(Typically, elastic scatterings with the plasma particles still  
continue after this period, so that kinetic equilibrium is maintained
down to lower temperatures, cf.\ e.g.\ ref.~\cite{kinetic}.)
A freeze-out process leading to chemical non-equilibrium  
is also responsible for the $\sim 20\%$ primordial helium abundance
observed in the Universe today, cf.\ e.g.\ ref.~\cite{bs0}. 

Analogous processes 
are assumed to play a role in heavy ion collisions. 
In particular, for $ \pi T \ll M $, 
the kinetic equilibration rate of heavy quarks
scales as $\Gamma_\rmi{kin} \sim \alpha_s^2 
\ln({\alpha_s}) T^2/M$~\cite{bs}--\cite{chm}, 
whereas the chemical equilibration rate scales as 
$\Gamma_\rmi{chem} \sim 
\alpha_s^2 T^{\fr32}\exp({-{M}/{T}}) /M^{\fr12} $~\cite{s1,s2}. 
Experimental data from RHIC and LHC suggest that 
charm quarks do have time to kinetically equilibrate, thereby 
participating in hydrodynamic flow (cf.\ e.g.\ refs.~\cite{rhic,lhc}), 
and theoretical efforts to understand this up to the non-perturbative
level are under way~\cite{hbm}--\cite{mumbai}. Building on earlier 
studies of strange quarks~\cite{msmc} it is believed, 
in contrast, that chemical 
equilibration does {\em not} take place; the number
density of charm quarks and antiquarks is essentially assumed to remain 
as determined by an initial hard process~\cite{abrs}, implying
that there are {\em more} heavy quarks present than would be 
due for chemical equilibrium (cf.\ e.g.\ ref.~\cite{chem}).

The purpose of this study is to  
suggest a definition of a chemical equilibration rate
of heavy quarks near equilibrium, 
similarly to what was achieved for their kinetic equilibration rate 
earlier on~\cite{cst,eucl}. 
A definition should be possible in the heavy-quark limit 
$ M \gg \pi T $, in which the rate itself is much slower than 
typical ``fast'' plasma rates, $\Gamma_\rmi{fast} \sim \alpha_s^n T$, 
$n \ge 1$. (If no scale separation is present between
$M$ and $\pi T$, 
then pair creations and annihilations take place as fast as 
elastic processes, and  
the massive degrees of freedom are to a good
approximation  in full thermal 
equilibrium with the strongly interacting heat bath.) 

The plan of this paper is the following. 
After some general considerations in \se\ref{se:general}, 
we recall the derivation of the chemical equilibration 
rate to leading order in $\alpha_s$, making use of the Boltzmann equation,   
in \se\ref{se:heur}. This is followed by a reminder that loop
corrections are likely to be substantial at any realistic
temperature, in \se\ref{se:loops}. A non-perturbative
formulation is put forward in \se\ref{se:nonpert}. Subsequently
we argue, in \se\ref{se:pert}, that in the weak-coupling 
limit the expression of \se\ref{se:nonpert} reduces to the 
result of \se\ref{se:heur}. 
A brief discussion of implications as well as prospects for 
non-perturbative studies 
concludes this writeup in \se\ref{se:concl}. 

%
\section{General considerations}
\la{se:general}

Assume that the system possess
an approximately conserved particle number.  Let us denote the
corresponding number density\footnote{%
  It is important to consider the number density rather than the
  differential phase space distribution, because otherwise it would be
  difficult to distinguish between processes changing the kinetic and
  the chemical decomposition of the system.  }  by $ n(t) $.  In
thermal equilibrium the value of $ n $ fluctuates around its
equilibrium value.  To treat the non-equilibrium problem we follow the
general method described in ref.~\cite{landau5}.  Let $\delta n(t)
\equiv n(t) - n_\rmi{eq}$ at some time $ t $ be large compared to the
mean fluctuation.  It will then evolve towards its equilibrium value.
Let us assume that the characteristic time scale $ \tau $ for this
evolution is much larger than the other relaxation times of the
system.  We only want to resolve time scales of order $ \tau $. Then
the non-equilibrium state is completely characterized by the
instantaneous value of $ \delta n $. Therefore the time derivative of $
\delta n $ can only depend on the value of $ \delta n $ and on 
thermodynamic quantities of the system such as temperature and
chemical potentials.  When $ \delta n $ is sufficiently small, one can
expand $ \delta \dot { n } $ in powers of $ \delta n $ and keep only
the linear term,
\begin{align} 
 \delta \dot{n}(t) = -\Gamma^{ }_\rmi{chem} \delta n(t) 
 \;. \la{clas_Lan}
\end{align} 
The coefficient $ \Gamma^{ }_\rmi{chem}  $ 
only depends on  thermodynamic quantities. 

Let us now be specific and choose $ n $ to be the sum of quark and
antiquark number densities, 
\begin{align}
  n \equiv  n^{ }_\rmii{$ Q $} + n^{ }_\rmii{$ \overline{Q} $}
  \;.
\end{align}   
We consider the heavy quark baryon number density 
 $  n^{ }_\rmii{$Q$} - n^{ }_\rmii{$ \overline{Q}$}$ 
to vanish (i.e.\ the baryon chemical potential to be zero).
We are interested in the limit that $ \pi T \ll M $. 
For heavy particles, 
$
 \{ \delta \dot{ n }(t)\}_\rmi{loss} \sim e^{-2 M/T}
$, 
because a heavy quark-antiquark pair gets annihilated,  
and 
$
 \delta n(t) \sim n_\rmi{eq} \sim e^{- M/T}
$.
Therefore $\Gamma^{ }_\rmi{}$ 
itself scales as $\sim e^{- M/T}$, implying
that this rate is much slower than most other processes in the system. In
particular, this rate
is slower than the kinetic equilibration rate. 
Therefore the heavy
quarks can be considered to be in kinetic equilibrium, which means that they
move very slowly. These almost static quarks experience rare number changing
reactions, and a non-perturbative description 
of the resulting dynamics, incorporating both the non-equilibrium
evolution of \eq\nr{clas_Lan} as well as equilibrium fluctuations, 
is presented in \eqs\nr{Lan1}--\nr{Gam_Om} below. 

%
\section{Boltzmann equation}
\la{se:heur}

If the system is weakly coupled, one can usually compute 
the coefficient $ \Gamma^{ }_\rmi{chem} $
in \eq(\ref{clas_Lan}), at least to leading
order, from the Boltzmann equation.
If we take into account $ 2 \to 2 $ scattering processes
and consider the limit 
$ \pi T \ll M $, it takes the 
form (cf.\ e.g.\ ref.~\cite{bbf}) 
\begin{align} 
    \dot n = - c \left ( n ^ 2 - n _ {\rm eq } ^ 2 \right ) 
  \equiv  \dot n  _ {\rm loss } +   \dot n  _ {\rm gain } 
  \label{boltzmann}
  \;, 
\end{align}
where  $\dot{n}_\rmi{loss} \equiv - c\, n^2$.
In equilibrium, with $n(t) \equiv n_\rmi{eq}$, gain and loss terms
must cancel  each other, and the number density is constant. Now linearize
(\ref{boltzmann}) as described in \se\ref{se:general}, which gives 
$ \delta  \dot n = - 2 c\, n  \delta  n $. Thus we can obtain 
$ \Gamma^{ }_\rmi{chem}  $ from the loss term in \eq(\ref{boltzmann}) via
\be
 \Gamma^{ }_\rmi{chem} = 
  - 2\, \frac{ \dot n _ {\rm loss } } { n _ {\rm eq } } 
 \;. \la{Gamma_heur_def}
\ee
An analogous discussion, implemented
by introducing separate ``chemical potentials'' for the quarks 
and antiquarks, can be found in ref.~\cite{msmc}.

Now we compute $ \Gamma^{ }_\rmi{chem}  $ 
using eq.~(\ref{Gamma_heur_def}) 
with tree-level matrix elements. 
The relevant loss processes are shown 
in \fig\ref{fig:heur}. Inserting the number of degrees of freedom
of the initial state, $2\Nc$, the decay rate
according to \eq\nr{Gamma_heur_def} can be written as
\ba
 \Gamma^{ }_\rmi{chem} 
 & = & \frac{2}{2\Nc \int_\vec{k} \nF{}(E_k)}
 \int 
 \prod_{a=1}^2
 \frac{{\rm d}^3 \vec{k}_a}{(2\pi)^3 2 E_{{k}_a} }
 \prod_{i=1}^2 \frac{{\rm d}^3 \vec{p}_i}{(2\pi)^3 2 \epsilon_{{p}_i}}
 \; (2\pi)^4 \delta^{(4)} (\mathcal{P}_1 + \mathcal{P}_2
 - \mathcal{K}_1  - \mathcal{K}_2 ) 
 \nn 
 & \times &
  \; \nF{}(E_{k_1}) \nF{}(E_{k_2})
  \biggl\{  
  \fr12 \nsum |\mathcal{M}^{ }_1|^2
  \, \bigl[ 1 + \nB{}(\epsilon_{p_1}) \bigr]
  \, \bigl[ 1 + \nB{}(\epsilon_{p_2}) \bigr]
 \nn & & \hspace*{2.6cm}
  + \, \Nf \nsum |\mathcal{M}^{ }_2|^2 
  \, \bigl[ 1 - \nF{}(\epsilon_{p_1}) \bigr]
  \, \bigl[ 1 - \nF{}(\epsilon_{p_2}) \bigr]
  \biggr\}
  \;. \la{Gamma_heur_1}
\ea
Here 
$
 \int_\vec{k} \equiv \int\! \frac{{\rm d}^3\vec{k}}{(2\pi)^3}
$; 
$\vec{k}_a$ are momenta in the initial state
and $\vec{p}_i$ those in the final state; 
$E_{k_a} \equiv \sqrt{k_a^2 + M^2}$ is the energy of a massive particle 
and $\epsilon_{p_i} \equiv |\vec{p}_i|$ is that of 
a massless one; and 
$\nF{}, \nB{}$ are the Fermi and Bose distributions, respectively.  
The sums are taken over the quantum numbers of 
all on-shell degrees of freedom, i.e.\ $2\Nc$ for quarks and antiquarks, 
and $2 d_A$ for gluons, with $d_A \equiv \Nc^2 - 1$. By $\Nf$
we denote the number of light quark flavours, and later on 
$C_F \equiv d_A / (2\Nc)$ will also appear. The factor $\fr12$
in front of the gluonic amplitude accounts for 
the two final state particles
being identical~\cite{msmc}.

%
\begin{figure}[t]
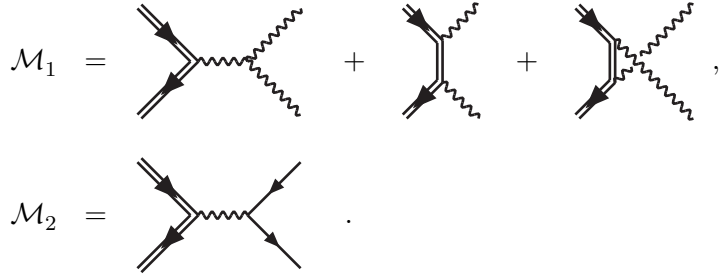


\hspace*{1.5cm}%
\begin{minipage}[c]{10cm}
\begin{eqnarray*}
 \mathcal{M}^{ }_1 
 & = & 
 \ScatA \quad + 
 \ScatB \hspace*{-5mm} +  
 \ScatC \hspace*{-2mm} \;, 
\\[5mm] 
 \mathcal{M}^{ }_2 
 & = & 
 \ScatD \quad\;. 
\end{eqnarray*}
\end{minipage}

\caption[a]{\small 
Scatterings through which an overabundance of heavy quarks
can disappear, assuming that there is an exponentially small 
thermal distribution of antiquarks present (or vica versa). 
A double line indicates heavy quarks, a single line light 
quarks, and a wiggly line gluons.}
\la{fig:heur}
\end{figure}
%

Taking the amplitude $\mathcal{M}^{ }_2$ of \fig\ref{fig:heur}
as an example, a text-book calculation yields
(cf.\ e.g.\ refs.~\cite{gor,pdg})
\be
 \nsum |\mathcal{M}^{ }_2|^2 = 
 \frac{4 g^4 C_F \Nc}{s^2}
 \Bigl[
  (M^2 -t)^2 + (M^2 - u)^2 + 2 M^2 s 
 \Bigr] 
 \;, \la{M2aver}
\ee
where $s,t,u$ are the standard kinematic invariants: 
$s \equiv (\mathcal{P}_1 + \mathcal{P}_2)^2 = 
(\mathcal{K}_1 + \mathcal{K}_2)^2$; 
$t \equiv (\mathcal{P}_1 -  \mathcal{K}_1)^2 = 
(\mathcal{P}_2 - \mathcal{K}_2)^2$; and
$u \equiv (\mathcal{P}_1 - \mathcal{K}_2)^2 = 
(\mathcal{P}_2 - \mathcal{K}_1)^2$.

The result simplifies further in the heavy-quark limit. 
Because of Boltzmann suppression
of $\nF{}(E_{k_a})$ at $M \gg \pi T$, we can consider the decaying
heavy quark and antiquark to be  almost at rest with respect
to the thermal medium: 
\be
 \mathcal{K}_1 \approx 
 \Bigl( M + \frac{k_1^2}{2M},\vec{k}_1 \Bigr)
 \;, \quad
 \mathcal{K}_2 \approx 
 \Bigl( M + \frac{k_2^2}{2M},\vec{k}_2 \Bigr)
 \;, \la{Boltzmann}
\ee
with $k_a \sim \sqrt{\pi T M} \ll M$. 
In contrast $p_1$ and $p_2$ are large because they 
have to carry away the energy liberated in the 
pair annihilation. So $\vec{k}_1 + \vec{k}_2$ can be 
approximated as zero in the phase
space constraints, and the Fermi distributions 
$\nF{}(\epsilon_{p_i})$ can be omitted: 
\ba
 \Gamma^{(q\bar{q})}_\rmi{chem} 
 & \approx & \frac{e^{- M/T}}{4 \Nc M^2}
 \int 
 \frac{{\rm d}^3 \vec{k}_2}{(2\pi)^3} \, e^{ -\frac{k_2^2}{2MT} }
 \nn & \times & 
 \frac{1}{(2\pi)^2} \int \!
 \frac{{\rm d}^3 \vec{p}_1}{2 \epsilon_{{p}_1}}
 \int \!
 \frac{{\rm d}^3 \vec{p}_2}{2 \epsilon_{{p}_2}}
 \; \delta^{(3)} (\vec{p}_1 + \vec{p}_2) 
 \delta (\epsilon_{p_1} + \epsilon_{p_2} - 2 M )
  \, \Nf \nsum |\mathcal{M}^{ }_2|^2 
  \;.
\ea
Here we cancelled a factorized integral against the one 
in the denominator. Noting also that 
\be
 s \approx 4 M^2
 \;, \quad 
 t \approx - M^2 
 \;, \quad
 u \approx - M^2  
 \;, \quad
 \la{massive}
\ee 
we get
$
 \sum |\mathcal{M}^{ }_2|^2 \approx
 4 g^4 C_F \Nc 
$.
The remaining integrals are trivially carried out, and we obtain
\be
 \Gamma^{(q\bar{q})}_\rmi{chem} \approx 
 \frac{g^4 C_F \Nf}{8\pi M^2}
 \Bigl( \frac{TM}{2\pi} \Bigr)^{\fr32}
 e^{-M/T}
 \;. \la{Gamma_heur_f}
\ee

A similar computation can be carried out with gluons, 
represented by the amplitude $\mathcal{M}^{ }_1$ of \fig\ref{fig:heur}.
Again the result is well-known (cf.\ e.g.\ refs.~\cite{gor,pdg}),
and reads 
\ba
 \nsum |\mathcal{M}^{ }_1|^2 
 \; = \; 
 4 g^4 C_F \Nc \biggl\{ 
 \!\!\! && \!\!\!
 4\Nc\, \frac{(M^2-t)(M^2-u)}{s^2}  
 \; + \; 
 (2 C_F - \Nc ) 
 \frac{2M^2(s-4M^2)}{(M^2 - t)(M^2 - u)}
 \nn 
 \!\!\! & + & \!\!\!
 2 C_F \biggl[ 
   \frac{(M^2-t)(M^2-u) - 2 M^2 (M^2+t)}{(M^2-t)^2}
   \; + (t\leftrightarrow u) 
 \biggr]
 \nn 
 \!\!\! & - & \!\!\!
 2 \Nc 
 \biggl[ 
   \frac{(M^2 - t)(M^2 - u) + M^2(u-t)}{s(M^2 - t)}
   \; + (t\leftrightarrow u) 
 \biggr] 
 \biggr\} 
 \;. \la{M2}
\ea
In the heavy-quark limit, \eq\nr{massive}, this simplifies to 
$
 \sum |\mathcal{M}^{ }_1|^2 \approx
 4 g^4 C_F \Nc (4 C_F - \Nc)
$.
The phase space integration goes through as before, and recalling
the $\fr12$ in \eq\nr{Gamma_heur_1}, 
\eq\nr{Gamma_heur_f} gets completed into 
\be
 \Gamma^{ }_\rmi{chem} \approx 
 \frac{g^4 C_F}{8\pi M^2}
 \Bigl( \Nf + 2 C_F - \fr{\Nc}2 \Bigr)
 \Bigl( \frac{TM}{2\pi} \Bigr)^{\fr32}
 e^{-M/T}
 \;. \la{Gamma_heur}
\ee
Numerically $2 C_F - \Nc/2 = \fr76$ for $\Nc = 3$; 
for $\Nf = 0$ this agrees with \eq(10) of ref.~\cite{s2}.
(We note, however, that for three light flavours, i.e.\ $\Nf = 3$, 
fermionic final states are significantly more 
important than purely gluonic ones.) 

%
\section{Towards loop corrections}
\la{se:loops} 

The result of \eq\nr{Gamma_heur} could well suffer from 
large radiative corrections. A few representative
examples of next-to-leading order (NLO) 
amplitudes are shown in \fig\ref{fig:loops}. In particular, the 
first amplitude, iterated by further rungs connecting the heavy 
quark and antiquark to each other, is responsible for binding the particles
to a quarkonium-like resonance. In the context of Dark Matter
co-annihilation, such a threshold enhancement is assumed to 
play a potentially important role, 
cf.\ e.g.\ refs.~\cite{gg,md}. However, 
this is not the only class of processes
in our case: as illustrated in \fig\ref{fig:loops}, all 
participating particles carry a colour charge, so that there 
may also be final-state interactions, as well as 
``non-factorizable'' terms 
connecting the initial and final states.

%
\begin{figure}[t]

\hspace*{1.5cm}%
\begin{minipage}[c]{13cm}
\begin{eqnarray*}
 \delta \mathcal{M}^{ }_2 
 & = & 
 \ScatDl  \quad + \quad 
 \ScatDk  \quad + \quad 
 \ScatDm  \quad + \ldots \;. 
\end{eqnarray*}
\end{minipage}

\caption[a]{\small 
Examples of 1-loop corrections to the scattering amplitude
$\mathcal{M}^{ }_2$ of \fig\ref{fig:heur}. }
\la{fig:loops}
\end{figure}
%

For future reference, we remark that there is one Euclidean
observable in which rungs between the heavy particles can also 
appear but which is nevertheless very well understood. This is 
the heavy quark-number susceptibility, formally defined as
\be
 \chi^{ }_{\ff}
 \equiv  \int_{\vec{x}}
 \Bigl\langle 
   (\bar\psi \gamma_0 \psi )(\tau,\vec{x})
   (\bar\psi \gamma_0 \psi )(0,\vec{0})
 \Bigr\rangle^{ }_T
 \;, \quad  0 \le \tau \le \beta
 \;, \quad \beta\equiv \frac{1}{T}
 \;. 
 \la{chi00}
\ee  
Because of charge conservation the argument $\tau$ can be chosen
at will. With vanishing chemical potentials, the susceptibility 
measures the mean number of heavy particles created by thermal
fluctuations, and is therefore closely related to the distribution
function $\nF{}(E_{k_2})$ on which the heavy quarks scatter
in \eq\nr{Gamma_heur_1}.

We recall that in the free limit the susceptibility evaluates to 
\be
 \chi^{ }_{\ff}
 = 
 4 \Nc \int \! \frac{{\rm d}^3\vec{k}}{(2\pi)^3}
 \, \nF{}(E_k)
 \, \bigl[ 1 - \nF{}(E_k) \bigr] 
 \;.   
\ee
For massless quarks the integral 
can be carried out in a closed form, yielding 
$ \chi^{ }_{\ff} =  \Nc T^3 / 3$, to which 
loop corrections are known up to a high order~\cite{av}, 
generically decreasing the susceptibility from the free value.
To us more relevant is the non-relativistic limit, 
\be
 \chi^{ }_{\ff} \approx 
 4 \Nc \int \! \frac{{\rm d}^3\vec{k}}{(2\pi)^3} e^{-E_k / T}
 \approx  
 4 \Nc\, \Bigl( \frac{MT}{2\pi} \Bigr)^{\fr32} e^{- M / T}
 \;. \la{chiff}
\ee 
Here the temperature dependence is precisely the same
as that in \eq\nr{Gamma_heur_f}.
Lattice data indicate that the susceptibility grows
rapidly with the temperature and, in the charm case, overcomes 
the exponential suppression already at temperatures of 
a few hundred MeV~\cite{milc}--\cite{buwu}, in line with the general 
expectation~\cite{pheneos}. We will keep these observations
in mind when estimating the numerical importance of 
the exponential suppression in \se\ref{se:concl}.

%
\section{Non-perturbative formulation} 
\la{se:nonpert}

Motivated by the remarks in \se\ref{se:loops}, the goal now 
is to suggest a non-perturbative {\em definition} of the heavy 
quark chemical equilibration rate. This could allow for
a systematic computation of higher order corrections, or in principle
be subjected e.g.\ to a lattice investigation.

In relativistic theories there is no obvious definition for a particle
number operator. Here we are interested in heavy quarks and antiquarks
with very small velocities.  In this case the energy of quarks and
antiquarks is roughly given by the sum of their rest energies or, in
other words, by their number density times the heavy quark mass $ M $.
Therefore the energy density of heavy quarks and antiquarks is a good
measure for their number density. We propose to define the
relaxation time of the number density 
$ n = n^{ }_\rmii{$Q$} + n^{ }_\rmii{$ \overline{Q} $}$
through the real time correlation function of the heavy quark Hamilton
operator.

We start by introducing an operator
describing heavy quark energy loss, both through elastic and through 
inelastic processes (\se\ref{ss:energy}); 
define then a ``transport coefficient'' 
related to this operator, capturing the desired rate (\se\ref{ss:transport}); 
and finally simplify one of the correlators appearing by considering
the heavy-quark limit (\se\ref{ss:heavy}).

%
\subsection{Operator for heavy quark energy loss} 
\la{ss:energy}

A form of the fermionic energy-momentum tensor 
which is symmetric, gauge-invariant, and leads to a correct finite
trace anomaly, reads~\cite{jcc1,jcc2}
\be
 T^{\mu\nu}_\rmi{f}
 \; \equiv \; 
 \frac{i}{4} \bar{\psi} 
 \Bigl(
  \gamma^\mu \overleftrightarrow{\!{D}}^{\!\nu}
 + 
  \gamma^\nu \overleftrightarrow{\!{D}}^{\!\mu}
 \Bigr) 
 \psi 
 - \eta^{\mu\nu} \, \mathcal{L}_\rmi{f} 
 \;. 
\ee
Here $\eta^{\mu\nu} \equiv \mathop{\mbox{diag}}$($+$$-$$-$$-$) and 
\be
 \bar{\psi} 
  \gamma^\mu \overleftrightarrow{\!{D}}^{\!\nu}
 \psi
 \; \equiv \;
 \bar{\psi} 
  \gamma^\mu \overrightarrow{\!{D}}^{\!\nu}
 \psi
 - 
 \bar{\psi} 
  \gamma^\mu \overleftarrow{\!{D}}^{\!\nu\dagger}
 \psi
 \;,
\ee
with 
$ 
 \overrightarrow{\!{D}}^{\!\nu}
 \psi 
 \equiv (\partial^\nu - i g A^\nu) \psi
$, 
$ 
 \bar{\psi}\,\overleftarrow{\!{D}}^{\!\nu\dagger}
 \equiv
 \bar{\psi}
 (\overleftarrow{\!{\partial}}^{\!\nu} + i g A^\nu) 
$, 
and $g$ denoting the bare gauge coupling. 
The Lagrangian can be written with a similar notation as 
\be
 \mathcal{L}_\rmi{f} = 
 \bar{\psi}
 \Bigl( 
   \frac{i}{2} \overleftrightarrow{\!{\bsl{D}}} - M 
 \Bigr)
 \psi
 \;. 
\ee

The heavy quark Hamilton operator is now defined by taking a spatial integral
over $T^{00}_\rmi{f}$, with the fields promoted to operators:  
\be
 \hat{H} \equiv \int_\vec{x} \hat{T}^{00}_\rmi{f}
 \; = \; 
 \int_\vec{x}
 \hat{\bar{\psi}}
 \Bigl( 
  - \frac{i}{2} \gamma^j \overleftrightarrow{\!{D}}_{\!j} + M 
 \Bigr)
 \hat{\psi}
  \;. \la{H}
\ee
Summation over repeated spatial indices is understood. 
Obviously, $\hat{H}$ could be written in other forms
by use of the Dirac equation, but for us it appears to be 
beneficial to employ a version with spatial derivatives only, 
because then partial integrations are formally allowed. 

In order to derive the operator for energy loss, let us 
also write down the Dirac equation in an explicit form, by 
placing time derivatives on the left-hand side: 
\ba
 \partial_t \hat{\psi} & = & 
 \Bigl[ 
   -i (M \gamma^0 - g A_0) - \gamma^0 \gamma^j 
    \overrightarrow{\!{D}}_{\!j}
 \Bigr] \hat{\psi}
 \;, \la{eom1} \\
 \partial_t \hat{\bar{\psi}} & = & 
 \hat{\bar{\psi}}
 \Bigl[ 
    i (M \gamma^0 - g A_0) - 
    \overleftarrow{\!{D}}_{\!j}^\dagger
    \gamma^j \gamma^0 
 \Bigr] 
 \;. \la{eom2}
\ea
In all of what follows,  
equations of motion are used for fermions only; 
derivatives acting on gauge fields are left ``as is'', 
formally assuming that gauge fields form a differentiable off-shell
background over which a path integral is to be carried out at a later stage. 

The task now is to construct $\partial_t \hat H$. The derivative
can act on any of the three possible locations in \eq\nr{H}:
\be
 \partial_t \hat{H} = 
 \int_\vec{x} 
  \biggl\{
    \Bigl(\partial_t \hat{\bar{\psi}} \Bigr)
      \Bigl( -i \gamma^j  \overrightarrow{\!{D}}_{\!j}  + M \Bigr) 
    \hat{\psi}
    + 
    \hat{\bar{\psi}} 
      \Bigl( -g \gamma^j \partial_0 A_j \Bigr) 
    \hat{\psi}
    + 
    \hat{\bar{\psi}} 
      \Bigl( i \gamma^j  \overleftarrow{\!{D}}_{\!j}^\dagger  + M \Bigr) 
    \Bigl( \partial_t    \hat{\psi}  \Bigr)   
  \biggr\}
 \;.
\ee
Inserting \eqs\nr{eom1}, \nr{eom2} and carrying out one partial 
integration, numerous cancellations take place, 
and we are finally left with 
\be
 \partial_t \hat{H}
 \; =  \;
 -g \int_\vec{x} \hat{\bar{\psi}} \gamma^j 
 \Bigl( 
   \partial_0 A_j - \partial_j A_0 - i g A_0 A_j + i g A_j A_0
 \Bigr)
 \hat{\psi}
 \; = \; 
 -g \int_\vec{x} \hat{\bar{\psi}}\, \gamma^j F^{ }_{0j}
 \hat{\psi}
 \;. \la{dtH}
\ee
So, in the presence of interactions ($g\neq 0$), the energy
carried by heavy quarks is not conserved. 

It appears that \eq\nr{dtH} has a classical interpretation. 
If a charged particle feels a Lorentz force, 
\be
 \frac{{\rm d}\vec{p}}{{\rm d}t} = q 
 \Bigl( \vec{E} + \vec{v}\times\vec{B} \Bigr)
 \;, 
\ee
then its energy changes as 
\be
 \frac{{\rm d}E}{{\rm d}t} = 
 \nabla_\vec{p}E \cdot  \frac{{\rm d}\vec{p}}{{\rm d}t} = 
 \vec{v} \cdot  \frac{{\rm d}\vec{p}}{{\rm d}t} = 
 q \vec{v}\cdot \vec{E}
 \;. 
\ee
Recalling that
$
 \hat{\bar{\psi}} \gamma^j \hat{\psi} 
$
are the spatial components of a current this is 
seen to agree in form with \eq\nr{dtH}. However, being a Fock space
operator, $\partial_t \hat{H}$ of \eq\nr{dtH}
describes also number-changing
reactions; in particular, if the initial state has more quarks and 
antiquarks than would be due for chemical equilibrium,  
a net pair annihilation 
should take place, 
and in the large-time limit
the corresponding matrix elements 
dominate the statistical average  of $\partial_t \hat{H}$.

%
\subsection{Defining a transport coefficient} 
\la{ss:transport}

To describe the depletion of
an overabundance of heavy quarks through a single coefficient, 
we follow a general method which has also been used for
determining their {\em kinetic equilibration rate}~\cite{cst,eucl}.
The goal is to relate the non-equilibrium rate
of interest, \eq\nr{clas_Lan}, 
to an equilibrium correlator, \eq\nr{Delta_c} (see ref.~\cite{landau5}
for a general argument concerning such relations). 
In order to achieve this goal, 
the logic is to use 
an ``effective'' classical picture to describe the long time physics of 
chemical equilibration. The  parameters of this description are 
subsequently matched to reproduce quantum-mechanical correlators. 
As we will see, the consistency of the description will 
be tested at the matching stage.

As discussed in \se\ref{se:general}, 
large deviations
from an equilibrium value tend to decrease, with a rate that
we want to determine 
(cf.\ \eq\nr{clas_Lan}); 
however, small deviations can also be 
generated by the occasional inverse reactions.\footnote{%
  In a heavy ion collision there may not be enough time 
  for inverse reactions to take place {\em in practice}; 
  but that does not change the {\em theoretical role} that they 
  play in relating 
  the non-equilibrium problem to a corresponding equilibrium one. 
  In other words, within the linear
  response regime the value of the coefficient $\Gamma_\rmii{chem}$ 
  is independent of initial conditions and of for how long we observe
  the dynamics.  
 } 
This is formally
the same physics as in Brownian motion, 
described by a Langevin equation, 
\ba
 \delta \dot{n}_{}(t) 
 &  = &   
 -\Gamma^{ }_\rmi{chem} \, \delta n_{}(t) + \xi (t) 
 \;, \la{Lan1} \\ 
 \langle\!\langle \, \xi (t) \, \xi (t') \, \rangle\!\rangle & = & 
  \Omega^{ }_\rmi{chem} \, \delta(t-t')
 \;, \qquad
 \langle\!\langle \xi(t) \rangle\!\rangle = 0 
 \;, \la{Lan2}
\ea
where $\delta n_{}$ is the non-equilibrium excess; 
$\xi$ is a 
stochastic noise, 
whose autocorrelation function is parametrized 
by $\Omega^{ }_\rmi{chem}$;  
and $\langle\!\langle ... \rangle\!\rangle$ denotes an average 
over the noise. The noise is uncorrelated because the time scale 
considered is much larger than any others in the system.\footnote{%
  At very short time scales, the noise is no longer white 
  but has a structure. By definition, the structure can be 
  resolved by inspecting the spectral function corresponding
  to the ``force-force'' correlator. As demonstrated 
  in \se\ref{se:pert}, the spectral function has support down to 
  small frequencies, with an overall magnitude 
  $\Omega_\rmii{chem}\sim e^{-2M/T}$. Noise becomes coloured at a frequency 
  scale $\omega^{ }_\rmii{UV}$ above which the shape  
  of the spectral function changes from its small-frequency asymptotics. 
  This is related to the physics of colour-electric fields, 
  so we may expect $\omega^{ }_\rmii{UV} \gsim \alpha_s^2 T$. This
  is much larger than the frequency scales that we are concerned 
  with, and plays no role in the following. 
}

Now, \eq\nr{Lan1} can be solved explicitly, given 
an initial value $\delta n(t_0)$:
\be
 \delta n_{}(t) = \delta n_{}(t_0)\, e^{-\Gamma_\rmii{chem} (t-t_0)}
 + 
 \int_{t_0}^t \! {\rm d}t' \, e^{\Gamma_\rmii{chem}(t'-t)} \xi(t')
 \;. \la{Lan_soln} 
\ee
Making use of this solution and taking an average 
over the noise, we can determine the 2-point 
correlation function of unequal time fluctuations of $\delta n_{}$:
\ba
 \Delta_\rmi{cl}(t,t')
 & \equiv & 
 \lim_{t_0 \to -\infty} 
 \langle\!\langle \, \delta n_{} (t) 
 \, \delta n_{} (t') \, \rangle\!\rangle
 \nn & = & 
 \lim_{t_0 \to -\infty} 
 \int_{t_0}^t \! {\rm d}t_1 \, e^{\Gamma_\rmii{chem}(t_1-t)} 
 \int_{t_0}^{t'} \! {\rm d}t_2 \, e^{\Gamma_\rmii{chem}(t_2-t')} 
 \langle\!\langle \, \xi(t_1)
 \, \xi (t_2) \, \rangle\!\rangle
 \nn & = & 
 \Omega^{ }_\rmi{chem} \lim_{t_0 \to -\infty} 
 \int_{t_0}^t \! {\rm d}t_1 \, e^{\Gamma_\rmii{chem}(t_1-t)} 
 \int_{t_0}^{t'} \! {\rm d}t_2 \, e^{\Gamma_\rmii{chem}(t_2-t')} 
 \delta(t_1-t_2)
 \nn & = & \frac{\Omega^{ }_\rmii{chem}}{2\Gamma_\rmii{chem}} \, e^{-
   \Gamma_\rmii{chem} | t-t'| } \;. \la{Delta_c} 
\ea 
 The limit $t_0\to -\infty$ here guarantees that any initial
 transients have died out; therefore, $\Delta_\rmi{cl}$ is an {\em
   equilibrium} correlation function.  Subsequently, making use of
$
 \partial_t \partial_{t'} |t-t'| = -2 \delta(t-t') 
$, 
we obtain
\be
 \partial_t \partial_{t'} \Delta_\rmi{cl}(t,t')
 =  - \frac{\Omega^{ }_\rmii{chem} \Gamma^{ }_\rmii{chem}}{2}\,  
 e^{- \Gamma_\rmii{chem} | t-t'| }
 + \Omega^{ }_\rmi{chem} \, \delta(t-t')
 \;. \la{ff_c} 
\ee
Fourier transforming
\eqs\nr{Delta_c} and \nr{ff_c} leads to  
\ba
 \tilde \Delta_\rmi{cl}(\omega) 
 & \equiv &  \int_{-\infty}^{\infty} \! {\rm d}t \, e^{i \omega (t-t')} 
 \Delta_\rmi{cl}(t,t')
 \; = \; 
 \frac{\Omega^{ }_\rmii{chem}}{ \omega^2 + \Gamma_\rmii{chem}^2}
 \;, \la{F_Delta_c} \\ 
 \omega^2 \tilde \Delta_\rmi{cl}(\omega) 
 & = &  \int_{-\infty}^{\infty} \! {\rm d}t \, e^{i \omega (t-t')} 
 \partial_t \partial_{t'} \Delta_\rmi{cl}(t,t')
 \; = \; 
 \frac{\omega^2 \, \Omega^{ }_\rmii{chem}}{ \omega^2 + \Gamma_\rmii{chem}^2}
 \;. \la{F_ff_c} 
\ea
It is also useful to note that, setting the time arguments equal, 
we can define a ``susceptibility'' as
\be
 \langle (\delta n_{})^2 \rangle_\rmi{cl}
 \equiv 
 \lim_{t_0 \to -\infty} 
 \langle\!\langle \, \delta n_{} (t) 
 \, \delta n_{} (t) \, \rangle\!\rangle 
 = \frac{\Omega^{ }_\rmii{chem}}{2\Gamma_\rmii{chem}}
 \;, \la{susc_c_def}
\ee
where we made use of \eq\nr{Delta_c}. 

Combining \eqs\nr{F_Delta_c}--\nr{susc_c_def}, various strategies
can be envisaged for determining the quantity that we are interested in, 
namely the non-equilibrium rate $\Gamma^{ }_\rmi{chem}$. 
A particularly fruitful approach is to 
take \eqs\nr{F_ff_c}, \nr{susc_c_def} 
as starting points, obtaining
\ba
 \Omega^{ }_\rmi{chem} & = & 
 \lim_{\Gamma_\rmii{chem} \,\ll\, \omega \,\ll\, \omega_\rmii{UV}} 
 \omega^2 \tilde \Delta_\rmi{cl}(\omega) 
 \;, \la{get_Lambda_c} \\ 
 \Gamma^{ }_\rmi{chem}  & = & 
 \frac{\Omega^{ }_\rmii{chem}}{2 \langle (\delta n_{})^2 \rangle_\rmi{cl}}
 \;. \la{get_Gamma_c} \la{Gam_Om}
\ea
Here $\omega_\rmii{UV}$ is a frequency scale at which some 
microscopic 
physics which is not described by the effective classical picture
sets in, typically 
$\omega_\rmii{UV} \sim \alpha_s^2 T $, 
and it has been assumed (cf.\ sec.~\ref{se:general}) that 
$\Gamma^{ }_\rmi{chem}$ {\em is parametrically small
compared with $\omega_\rmii{UV}$}. In our case this is so because
$\Gamma^{ }_\rmi{chem}$ is exponentially suppressed 
as $\sim e^{- M / T}$. 
With this input, all real-time information is in the numerator
of the equilibrium correlator $\omega^2\tilde \Delta_\rmi{cl}(\omega)$. 

After these preparatory steps, we can promote the determination 
of $\Gamma^{ }_\rmi{chem}$ to the quantum level. 
It just remains to note that since 
in the classical limit observables commute, a suitable 
quantum version of the equilibrium correlator is 
\be
 \Delta_\rmi{qm}(t,t') \equiv  \Bigl\langle \fr12 
 \bigl\{ \delta \hat{n}_{}(t), \delta \hat{n}_{}(t') \bigr\} 
 \Bigr\rangle
 \;. \la{Delta_q} 
\ee
So, \eqs\nr{get_Lambda_c}, \nr{get_Gamma_c} can be rephrased as
\ba
 \Omega^{ }_\rmi{chem} & = & 
  \lim_{\Gamma_\rmii{chem} \,\ll\, \omega \,\ll\, \omega_\rmii{UV}} 
  \omega  ^ 2 
  \int_{-\infty}^{\infty} \! {\rm d}t \, e^{i \omega (t-t')} 
  \biggl\langle \fr12 
  \Bigl\{   \delta \hat{n}_{}(t), 
   \delta \hat{n}_{}(t') 
  \Bigr\} 
  \biggr\rangle
  \;, \la{get_Lambda_q} 
 \label{gammafromdelta}
\ea
or 
\ba
 \Omega^{ }_\rmi{chem} & = & 
  \lim_{\Gamma_\rmii{chem} \,\ll\, \omega \,\ll\, \omega_\rmii{UV}} 
  \int_{-\infty}^{\infty} \! {\rm d}t \, e^{i \omega (t-t')} 
  \biggl\langle \fr12 
  \Bigl\{ \frac{{\rm d}  \hat{n}_{}(t)}{{\rm d}t}, 
  \frac{{\rm d} \hat{n}_{}(t')}{{\rm d}t'} 
  \Bigr\} 
  \biggr\rangle
  \;, \la{get_Lambda_q_2} 
\ea
together with
\ba
 \Gamma^{ }_\rmi{chem}  & = & 
 \frac{\Omega^{ }_\rmii{chem}}{2 \langle (\delta \hat{n}_{})^2 \rangle^{ }}
 \;. \la{get_Gamma_q}
\ea
The denominator of \eq\nr{get_Gamma_q} 
is nothing but the variance, 
$
 \langle (\delta \hat{n}_{})^2 \rangle 
 = \langle \hat{n}_{}^2 \rangle^{ }_{ } - 
   \langle \hat{n}_{} \rangle^2_{ } 
$.
The consistency of the matching is tested 
at least to some extent by whether the variance 
is UV-finite (for most composite operators this is not the case).

The formulae introduced can be applied on a non-perturbative
level by re-expressing them  through the 
imaginary-time formalism. This means that we first define
a Euclidean correlator, $\Omega(\tau)$; 
Fourier-transform it, 
$
  \tilde \Omega (\omega_n) = 
  \int_0^\beta \! {\rm d}\tau \, e^{i \omega_n\tau } \Omega (\tau)
$, 
where $\omega_n = 2\pi n T$, $n \in \mathbbm{Z}$
(this requires the presence of an UV regulator, or 
the subtraction of short-distance divergences); and 
obtain the spectral function from its imaginary part, 
$
  \rho^{ }_\rmii{$\Omega$} (\omega) = 
  \im \tilde \Omega (\omega_n \to -i [\omega + i 0^+])
$.
The symmetric combination needed in \eq\nr{get_Lambda_q_2} 
is given by 
$
 \Omega^{ }_\rmi{chem} = 
 \lim^{ }_{\;\Gamma_\rmii{chem} \,\ll\, \omega \,\ll\, \omega_\rmii{UV}} 
 {2 T \rho^{ }_\rmii{$\Omega$}(\omega)} / {\omega}
$.


The argumentation above can directly be transported
to the case at hand, with $\hat n$ replaced by $\hat H$ from 
\eq\nr{H}.
Denoting by $E_j$ the {\em Euclidean} electric 
field, which contains an additional $i$ from a Wick rotation, the 
imaginary-time correlator referred to above reads
(we divide by volume in order to define intensive quantities)
\ba
 \Omega(\tau) & \equiv & 
 \frac{1}{V} \,
 \Bigl\langle \partial_t\hat{H}(\tau) \,  
              \partial_t\hat{H}(0) \Bigr\rangle_\rmi{qc} 
 \nn 
 & = & 
 -g^2 \int_\vec{x}
 \Bigl\langle 
 \bigl[\bar{\psi} \gamma^j E_j \psi \bigr] (\tau,\vec{x})
 \,
 \bigl[\bar{\psi} \gamma^k E_k \psi \bigr] (0,\vec{0})
 \Bigr\rangle_\rmi{qc} 
 \;, \la{Omega_tau}
\ea
where
$
 g E_k \equiv i [D_\tau,D_k]
$, 
and 
$\langle...\rangle_\rmi{qc}$
refers to connected quark contractions
(the reason for this choice is discussed  
in \fig\ref{fig:pert}).
Hats have been left out in the second row
because this correlator can be evaluated
with regular path integral techniques. 
Similarly, the correlator related to energy fluctuations becomes 
\ba 
 \Delta(\tau) & \equiv & 
 \frac{1}{V} \,
 \Bigl\langle \hat{H}(\tau) \,  
              \hat{H}(0) \Bigr\rangle_{\rmi{c}}
 \nn 
 & = & 
 \int_\vec{x}
 \Bigl\langle 
 \Bigl[\bar{\psi} 
  \Bigl(
      - \frac{i}{2} \gamma^j \overleftrightarrow{\!{D}}_{\!j} + M  
  \Bigr)
 \psi \Bigr] (\tau,\vec{x})
 \,
 \Bigl[\bar{\psi} 
  \Bigl(
      - \frac{i}{2} \gamma^k \overleftrightarrow{\!{D}}_{\!k} + M  
  \Bigr)
 \psi \Bigr] (0,\vec{0})
 \Bigr\rangle_{\rmi{c}}
 \;, \la{Phi_tau}
\ea
where $\langle...\rangle_\rmi{c}$ refers to the connected part, 
i.e.\ 
$\langle \hat H(\tau) \hat H(0) \rangle_\rmi{c} \equiv
\langle \hat H(\tau) \hat H(0) \rangle - 
\langle \hat H(0) \rangle^2
$. 
We can interpret $\Delta(\tau)$ as the susceptibility
needed in \eq\nr{get_Gamma_q} to the extent that it is 
$\tau$-independent and therefore finite at $\tau \to 0$
(cf.\ \eq\nr{chi00}); this turns out to be the case  
in the limit $\pi T \ll M$, where it  
corresponds to a quasi-conserved quantity:
$
 \Delta(\tau) \approx
 \frac{1}{V} \langle (\delta \hat{H}_{})^2 \rangle 
$ (cf.\ \eq\nr{Phi_tau_3}).

%
\subsection{Heavy quark limit} 
\la{ss:heavy}

The correlators in \eqs\nr{Omega_tau}, \nr{Phi_tau} can be
understood physically, and also written in somewhat simpler forms, 
if two-component spinors 
corresponding to non-relativistic degrees of freedom are employed.
We choose a representation for the Dirac matrices with
\be
 \gamma^0 \equiv
 \left( 
  \begin{array}{cc}
   \unit & \;0 \\   
   0 & -\unit 
  \end{array}
 \right)
 \;, \quad
 \gamma^k \equiv
 \left( 
  \begin{array}{cc}
   \;0 & \sigma_k \\   
   -\sigma_k & 0 
  \end{array}
 \right)
 \;, \quad k = 1,2,3
 \;, \la{Dirac_NRQCD}
\ee
where $\sigma_k$ are the Pauli matrices. The Dirac spinors are 
written as 
\be
 \psi \, \equiv  
  \left( 
  \begin{array}{c} 
    \theta \\ \chi
  \end{array}
 \right)
 \;, \quad
 \bar\psi \, \equiv \,
 ( \theta^\dagger \;, \; - \chi^\dagger ) 
 \;. \la{nr}
\ee
Clearly $\theta$ corresponds to $\mathbbm{P}^{ }_+ \psi$ and 
$\chi$ to $\mathbbm{P}^{ }_-\psi$, with the projection operators defined as
$
 \mathbbm{P}_\pm \; \equiv \; \fr12 \bigl( \mathbbm{1} \pm \gamma^0  \bigr)
$.
With this notation the operator entering \eq\nr{Omega_tau} can be expressed as
\be
 \partial_t H = 
 -i g
 \int_\vec{x}
 \bigl[
   \theta^\dagger \vec{\sigma}\cdot\vec{E}\, \chi +  
   \chi^\dagger \vec{\sigma}\cdot\vec{E}\, \theta
 \bigr]
 \;. \la{Omega_tau_2}
\ee
Note that this 
operator is {\em different} from that 
relevant for heavy quark {\em kinetic} equilibration: electric fields
appear in both cases but here they come together with 
$\theta^\dagger\chi$, $\chi^\dagger \theta$, whereas in 
ref.~\cite{eucl} the combinations 
$\theta^\dagger\theta$, 
$\chi^\dagger\chi$ appeared. 
Eq.~\nr{Phi_tau} can also be expressed in the new notation, 
with the Hamiltonian becoming
\be
 H = \int_\vec{x} \Bigl[ 
   M \bigl( \theta^\dagger\theta - \chi^\dagger\chi \bigr)
  - \frac{i}{2} \Bigl( 
   \theta^\dagger \vec{\sigma} \cdot 
   \overleftrightarrow{{\vec{D}}} \chi +
   \chi^\dagger \vec{\sigma} \cdot 
   \overleftrightarrow{{\vec{D}}} \theta 
  \Bigr)
 \Bigr]
 \;. \la{Phi_tau_2}
\ee

For a proper physical interpretation, it is useful to change
the ordering of $\chi^*_\alpha$, $\chi^{ }_\beta$. 
It then becomes clear that 
$\chi^*$ represents an antiparticle to $\theta$; a most 
direct way to see this is  from 
the number density operator: 
$
 \bar{\psi}\gamma^0\psi = 
 \bar{\psi}(P^{ }_+ - P^{ }_-)\psi = 
 \theta^\dagger\theta + \chi^\dagger \chi =
 \theta^\dagger\theta - {\chi^*}^\dagger \chi^* 
$. 
What this implies is that operators of the types $\theta^\dagger\chi$, 
$\chi^\dagger\theta$, appearing in \eq\nr{Omega_tau_2}, create or 
annihilate quark-antiquark pairs; and that the leading term 
of the Hamilton operator in \eq\nr{Phi_tau_2} counts particles
{\em plus} antiparticles, assigning each 
energies given by their rest mass. 

After these remarks we can simplify 
the correlator $\Delta(\tau)$ of \eq\nr{Phi_tau}. 
In the heavy-quark limit the leading term comes from  
$M(\theta^\dagger\theta - \chi^\dagger \chi)$ in \eq\nr{Phi_tau_2}.
But since in the same limit the cross term gives no contribution, 
the (disconnect part of) the 2-point correlator is the same 
as that for 
$\bar{\psi}\gamma^0\psi = \theta^\dagger\theta + \chi^\dagger \chi$.
So, 
\be
 \Delta(\tau) \; \approx \; 
 M^2 \chi^{ }_{\ff}  
 \; = \; 
 M^2  \int_{\vec{x}}
 \Bigl\langle 
   (\bar\psi \gamma_0 \psi )(\tau,\vec{x})
   (\bar\psi \gamma_0 \psi )(0,\vec{0})
 \Bigr\rangle^{ }_T
 \;, \la{Phi_tau_3}
\ee
where $\chi^{ }_{\ff}$ is from \eq\nr{chi00}.
As required, \eq\nr{Phi_tau_3} is independent of $\tau$.
Unfortunately, for $\Omega(\tau)$ of \eq\nr{Omega_tau}, 
it is not clear to us whether 
any similar simplification is possible; the reasons for this 
are discussed at the beginning of \se\ref{se:pert}.

To summarize, from the Euclidean correlator, $\Omega(\tau)$ 
in \eq\nr{Omega_tau}, we 
can in principle construct the 
Matsubara representation, 
$\tilde \Omega(\omega_n) \equiv \int_0^\beta \! {\rm d}\tau \, 
 e^{ i \omega_n\tau} \Omega(\tau)$, if an ultraviolet
regulator or subtraction is present.
After analytic continuation, 
$
 \rho^{ }_\rmii{$\Omega$}(\omega)
 = \im \Omega(\omega_n \to -i [\omega + i 0^+])
$, 
the decay rate of \eq\nr{get_Gamma_q} follows from 
\be
 \Gamma^{ }_\rmi{chem} 
 \equiv  \frac{\lim_{\omega \to 0^+} 
 \frac{\raise0.6ex\hbox{\small $2 T \rho^{ }_\rmii{$\Omega$}(\omega)$}}
 {\omega}}
 {2 \chi^{ }_{\ff} M^2}
 = 
 \lim_{\omega \to 0^+} 
 \biggl\{
  \frac{T \rho^{ }_\rmii{$\Omega$}(\omega)}
 {\omega \chi^{ }_{\ff} M^2 } 
 \biggr\}
 \;. \la{Gamma_compact}
\ee

We remark that since \eq\nr{Omega_tau} involves composite operators
for non-conserved quantities, 
the issue of renormalization is non-trivial. 
Unfortunately a satisfactory discussion  
goes beyond the scope of the present work. 

%
\section{Perturbative evaluation} 
\la{se:pert}

So far we have made no approximation based on
the weak-coupling expansion. At high $T$, 
however, the renormalized gauge coupling can be assumed small; 
we would like to make use of this limit in order to compare 
the general formulae with those in \se\ref{se:heur}. 

It is now important to be more precise about the nature
of the heavy-quark limit. Even though we made use of the
``non-relativistic'' spinors $\theta$ and $\chi$ in \se\ref{ss:heavy}
in order to obtain a physical interpretation for the operators
appearing, the function $\Omega(\tau)$ {\em cannot} actually be
evaluated with non-relativistic kinematics. A trivial reason
is that with non-relativistic dispersion relations, a heavy quark 
and antiquark can annihilate into a single gluon; this non-sensical
reaction would spoil the physics. In addition, in the $t$ and $u$-channel
processes of \fig\ref{fig:heur} the heavy quarks are deeply virtual, 
cf.\ \eq\nr{massive}. 
That said, some parts of the analysis can still be simplified, 
but {\em a priori} the quark propagators need to be fully relativistic. 

The relevant graphs are shown in \fig\ref{fig:pert}. It is
easy to see that the leading-order graph, (a), does not 
contribute: after analytic continuation and taking the cut
we are faced with the decay of a heavy quark and 
a heavy antiquark into a gluon, which is forbidden by
relativistic kinematics.
At NLO, in contrast, there are non-vanishing 
contributions;  let us show this explicitly by evaluating
the fermionic graph in \fig\ref{fig:eucl_details}.

To get started, we note that in its original form 
the amplitude squared of \eq\nr{M2aver} reads 
\ba
 \Nf \nsum |\mathcal{M}^{ }_2|^2 
 & = &  
      {g^4 \Nf \tr[T^a T^b] \tr[T^a T^b]}
 \nn[2mm] & & \times \, 
 \frac{
 \tr[\gamma^\mu \bsl{\mathcal{P}_1} \gamma^\nu \bsl{\mathcal{P}_2}]
 \tr[\gamma_\mu (\bsl{\mathcal{K}_1} +M )
      \gamma_\nu (\bsl{\mathcal{K}_2} -M )] }
 {(\mathcal{P}_1 + \mathcal{P}_2)^4}
 \;, \la{M2_orig}
\ea
where $T^a$ are the Hermitean generators of SU($\Nc$), 
normalized as $\tr[T^a T^b] = \frac{\delta^{ab}}{2}$; 
whereas the 
imaginary time
diagram of \fig\ref{fig:eucl_details} 
can be written as 
\ba
 \tilde\Omega^{(q\bar{q})}(\omega_n)
 & = & -g^4 \Nf \tr[T_{ }^a T_{ }^b]\, \tr[T_{ }^a T_{ }^b]\; 
 \nn[2mm] & & \times \, 
  \Tint{\{P_1 P_2 K_1 K_2\}}
 \hspace*{-1cm}
 \frac{ \deltabar(\omega_n+P_1+P_2-K_1-K_2)
 \, {\varepsilon_{\mu;\alpha}(P_1+P_2)
 \varepsilon_{\nu;\beta}(P_1+P_2)} 
 }
 {
  P_1^2 P_2^2(K_1^2 + M^2)(K_2^2 + M^2)
 }
 \nn[2mm] & & \times \,  
 \frac{\tr[\gamma_\alpha (i \bsl{{P}_1}) \gamma_\beta (i \bsl{{P}_2})]
 \tr[\gamma_\mu (i \bsl{{K}_1} +M )
      \gamma_\nu (i \bsl{{K}_2} -M )]
 }{(P_1+P_2)^4 }
 \;. 
 \la{M2_eucl}
\ea
Here four-momenta and Dirac-matrices are Euclidean; 
$\omega_n$ within the $\;\deltabar$ is a short-hand for
$(\omega_n,\vec{0})$; 
$\deltabar$ is normalized so that $\Tinti{P}\;\deltabar(P) = 1$; 
sum-integrals are standard, with $\Tinti{\{...\}}$ denoting fermionic 
Matsubara frequencies; and
\be
 \varepsilon_{\mu;\alpha}(P) \equiv 
 P_0 \,\delta_{\mu\alpha} - P_\mu\, \delta_{0\alpha} 
 \la{epsilon}
\ee
originates from the electric fields. 
A close kinship between \eqs\nr{M2_orig}, \nr{M2_eucl} is 
immediately observed, but to see that they really lead to 
the same physics requires a careful analysis.

%
\begin{figure}[t]
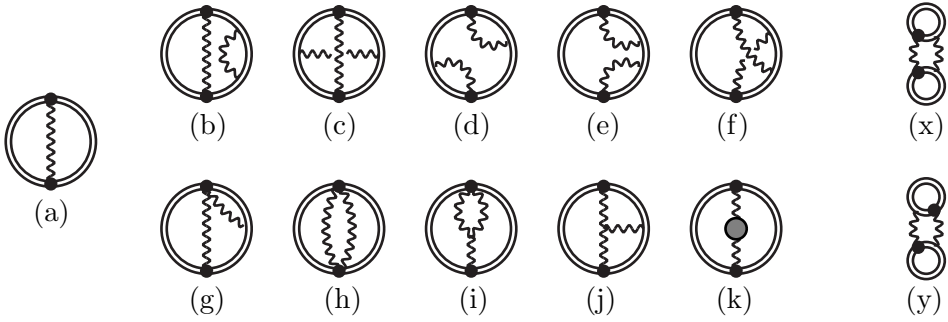


\hspace*{0.2cm}%
\begin{minipage}[c]{3cm}
\begin{eqnarray*}
&& 
 \hspace*{-1cm}
 \EleA 
\\[1mm] 
&& 
 \hspace*{-0.6cm}
 \mbox{(a)} 
\end{eqnarray*}
\end{minipage}%
\begin{minipage}[c]{11.2cm}
\begin{eqnarray*}
&& 
 \hspace*{-1cm}
 \EleB \quad\; 
 \EleBB \quad\; 
 \EleC \quad\; 
 \EleD \quad\; 
 \EleE \quad\qquad\; 
 \EleK \quad\; 
\\[1mm] 
&& 
 \hspace*{-0.6cm}
 \mbox{(b)} \hspace*{1.26cm}
 \mbox{(c)} \hspace*{1.26cm}
 \mbox{(d)} \hspace*{1.26cm}
 \mbox{(e)} \hspace*{1.26cm}
 \mbox{(f)} \hspace*{2.10cm}
 \mbox{(x)} 
\\[5mm] 
&& 
 \hspace*{-1cm}
 \EleF \quad\; 
 \EleG \quad\; 
 \EleH \quad\; 
 \EleI \quad\; 
 \EleJ \quad\qquad\; 
 \EleL \quad 
\\[1mm] 
&& 
 \hspace*{-0.6cm}
 \mbox{(g)} \hspace*{1.3cm}
 \mbox{(h)} \hspace*{1.28cm}
 \mbox{(i)} \hspace*{1.28cm}
 \mbox{(j)} \hspace*{1.32cm}
 \mbox{(k)} \hspace*{2.02cm}
 \mbox{(y)} 
\end{eqnarray*}
\end{minipage}

\caption[a]{\small 
The graphs contributing to the correlator 
$\Omega(\tau)$ defined in \eq\nr{Omega_tau}, 
up to $\rmO(g^4)$ (time runs vertically). 
The double lines denote heavy quarks; the small dots the composite 
operators; and the grey blob the 1-loop gauge field self-energy. 
Graphs (a)-(k) look similar to those relevant for computing 
the correlator yielding the heavy quark kinetic equilibration 
rate~\cite{rhoE}, but the kinematic regime is different. The additional
graphs (x) and (y) amount to a renormalization of the gluonic
part of the energy-momentum tensor by virtual heavy quarks, 
and have been excluded from the definition in  \eq\nr{Omega_tau}
by restricting to connected quark contractions.
} 
\la{fig:pert}
\end{figure}
%

%
\begin{figure}[t]
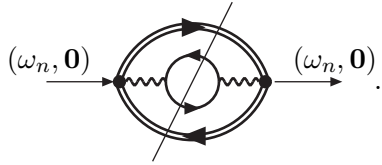


\hspace*{1.5cm}%
\begin{minipage}[c]{10cm}
\begin{eqnarray*}
 && \hspace*{20mm}
 \ScatDe  \quad\qquad\;. 
\end{eqnarray*}
\end{minipage}

\vspace*{3mm}

\caption[a]{\small 
The part of diagram (k) of \fig\ref{fig:pert} sensitive to light 
quarks, after a Fourier transformation to Euclidean frequency
$\omega_n$
and a rotation by 90 degrees. 
The diagonal line indicates a cut. } 
\la{fig:eucl_details}
\end{figure}
%

We note, first of all, that the index
$\mu$ appearing in \eq\nr{epsilon} can only be spatial. Therefore,  
in the heavy-quark part
\be
  \tr[\gamma_\mu (i \bsl{{K}_1} +M )
      \gamma_\nu (i \bsl{{K}_2} -M )] = 
  4 \bigl[
  \delta_{\mu\nu}(K_1\cdot K_2 - M^2) 
 - K_{1\mu} K_{2\nu} - K_{1\nu} K_{2\mu} 
    \bigr]
 \;,
\ee
we can drop the terms $- K_{1\mu} K_{2\nu} - K_{1\nu} K_{2\mu}$ and 
the spatial part of $K_1 \cdot K_2$, 
because the heavy quarks will 
be non-relativistic, cf.\ \eq\nr{Boltzmann}. The part containing
final-state momenta,  
\ba
 & & \hspace*{-1cm}
 \delta_{i\mu}\delta_{i\nu}
 \varepsilon_{\mu;\alpha}(P_1+P_2) \varepsilon_{\nu;\beta}(P_1+P_2)
 \tr[\gamma_\alpha (i \bsl{{P}_1}) \gamma_\beta (i \bsl{{P}_2})]
 \nn & = & 
 4 \,
 \varepsilon_{i;\alpha}(P_1+P_2) \varepsilon_{i;\beta}(P_1+P_2)
 \bigl[
  \delta_{\alpha\beta} P_1\cdot P_2  
 - P_{1\alpha} P_{2\beta} - P_{1\beta} P_{2\alpha} 
 \bigr]
 \;,
\ea
can in turn be re-expressed as ($P_i \equiv (p_{ni},\vec{p}_i)$)
\ba
 \varepsilon_{i;\alpha} \, \varepsilon_{i;\beta} \,
 \delta_{\alpha\beta} & = & 
 3 (P_1 + P_2)^2 - 2 (\vec{p}_1 + \vec{p}_2)^2  \;, 
 \\ 
 \varepsilon_{i;\alpha} \, \varepsilon_{i;\beta} \,
 P_{1\alpha} P_{2\beta} & = & 
 (P_1+P_2)^2 p_{n1}p_{n2} 
 - P_1^2 (p_{n1}+p_{n2})p_{n2} 
 - P_2^2 (p_{n1}+p_{n2})p_{n1}
 \;. \la{funny}
\ea
The latter two terms of \eq\nr{funny} do not 
contribute due to the antisymmetry in one of the summation 
variables (for instance, in the middle term, after first 
carrying out $T\sum_{p_{n1}}$ the expression is antisymmetric in $p_{n2}$), 
so we get 
\ba
 & & \hspace*{-1cm}
 \tilde\Omega^{(q\bar{q})}(\omega_n)
  \approx  - 8 g^4 C_F \Nc \Nf \; 
  \Tint{\{P_1 P_2 K_1 K_2\}}
 \hspace*{-1cm} \deltabar(\omega_n+P_1+P_2-K_1-K_2)
 \frac{
 k_{n1}k_{n2} - M^2
 }{(K_1^2 + M^2)(K_2^2 + M^2)}
 \nn & \times & 
 \frac{1}{P_1^2}\biggl\{ 
    \frac{3}{2 P_2^2}
    - \frac{3}{(P_1+P_2)^2} 
    + \frac{2 (\vec{p}_1 + \vec{p}_2)^2}{ (P_1+P_2)^4}
    - \frac{(\vec{p}_1 + \vec{p}_2)^2}{P_2^2 (P_1+P_2)^2}
    - \frac{2 p_{n1}p_{n2}} {P_2^2 (P_1+P_2)^2}
 \biggr\}
 \;.
 \la{M2_eucl_2}
\ea

To carry out the Matsubara sums, we write 
\be
 \delta(\omega_n + p_{n1} + p_{n2} - k_{n1} - k_{n2})
 = \int_0^\beta \! {\rm d}\tau \, 
 e^{i (\omega_n + p_{n1} + p_{n2} - k_{n1} - k_{n2})\tau}
 \;. 
\ee
Then,  
\be
 T^2 \!\!\!\! \sum_{ \{ k_{n1} k_{n2} \} } \!\!
 \frac{
 (k_{n1}k_{n2} - M^2)e^{-i (k_{n1} + k_{n2})\tau}
 }{(k_{n1}^2 + E_{k_1}^2)(k_{n2}^2 + E_{k_2}^2)}
  \approx  -\frac{1}{2}
 \frac{
  e^{(\beta-\tau)(E_{k_1} + E_{k_2})} 
  + 
  e^{\tau(E_{k_1} + E_{k_2})} 
 }{
 (e^{\beta E_{k_1}} + 1)
 (e^{\beta E_{k_2}} + 1)
 }
 \;, 
\ee
where we again approximated $E_{k_a} \approx M$ 
in the spin part (but not in the exponential functions), 
whereby the ``crossed terms'' cancelled in the sum. 
As far as the second row of \eq\nr{M2_eucl_2} is concerned, 
we note that in the 2nd and 3rd terms a shift $p_{n2}\to p_{n2}-p_{n1}$
factorizes the $p_{n1}$-dependence from the $\tau$-dependence. 
These terms lead to a vanishing contribution
to the transport coefficient defined in \eq\nr{Gamma_compact};
the reason is that since neither 
$\epsilon_{p_1}$ nor $\epsilon_{p_2}$ appears in the time 
dependence, we are 
left with the phase space constraints 
$
 \delta(E_{k_1} + E_{q k_1} - q)
$
or 
$
 \delta(E_{k_1} + E_{q k_1} + q)
$, 
where $E_{q k_1} \equiv \sqrt{(\vec{q - k_1})^2 + M^2}$
and $\vec{q} \equiv \vec{p}_1 + \vec{p}_2$. These constraints
cannot get realized and so the factorized terms can be 
omitted.\footnote{%
 In the case with the ``double pole'',
 i.e.\ the 3rd term of \eq\nr{M2_eucl_2},  
 one can replace 
 $(P_1+P_2)^2 \to (P_1+P_2)^2 + m_0^2$; 
 consider first a single pole; and take subsequently 
 a derivative with respect to $m_0^2$. The relevant phase space 
 constraint becomes 
 $
  \delta(E_{k_1} + E_{q k_1} - \epsilon_q)
 $, 
 with $\epsilon_q \equiv \sqrt{q^2 + m_0^2}$. This does not get realized
 if $m_0 < 2 M$, so the function vanishes exactly in this regime, 
 and thereby the derivative vanishes as well.}

Non-trivial contributions arise from the remaining three terms
of \eq\nr{M2_eucl_2}. Defining
\ba
 \tilde{\mathcal{I}}_1(\omega_n) \!\! & \equiv & \!\! 
 \int_0^\beta \! {\rm d}\tau \, 
 e^{i \omega_n \tau} 
 \, \frac{ 
  e^{(\beta-\tau)(E_{k_1} + E_{k_2})} 
  + 
  e^{\tau(E_{k_1} + E_{k_2})} 
 }{
 (e^{\beta E_{k_1}} + 1)
 (e^{\beta E_{k_2}} + 1)
 }
 \, 
 T^2 \!\!\!\! \sum_{ \{ p_{n1} p_{n2} \} }
 \frac{e^{i(p_{n1} + p_{n2})\tau}}{P_1^2 P_2^2}
 \;, \la{I1wn} \\ 
 \tilde{\mathcal{I}}_2(\omega_n) \!\! & \equiv & \!\! 
 \int_0^\beta \! {\rm d}\tau \, 
 e^{i \omega_n \tau} 
 \, \frac{ 
  e^{(\beta-\tau)(E_{k_1} + E_{k_2})} 
  + 
  e^{\tau(E_{k_1} + E_{k_2})} 
 }{
 (e^{\beta E_{k_1}} + 1)
 (e^{\beta E_{k_2}} + 1)
 }
 \, 
 T^2 \!\!\!\! \sum_{ \{ p_{n1} p_{n2} \} }
 \frac{e^{i(p_{n1} + p_{n2})\tau}(\vec{p}_1+\vec{p}_2)^2}
 {P_1^2 P_2^2 (P_1 + P_2)^2}
 \;, \nn \la{I2wn} \\ 
 \tilde{\mathcal{I}}_3(\omega_n) \!\! & \equiv & \!\! 
 \int_0^\beta \! {\rm d}\tau \, 
 e^{i \omega_n \tau} 
 \, \frac{ 
  e^{(\beta-\tau)(E_{k_1} + E_{k_2})} 
  + 
  e^{\tau(E_{k_1} + E_{k_2})} 
 }{
 (e^{\beta E_{k_1}} + 1)
 (e^{\beta E_{k_2}} + 1)
 }
 \, 
 T^2 \!\!\!\! \sum_{ \{ p_{n1} p_{n2} \} }
 \frac{e^{i(p_{n1} + p_{n2})\tau} p_{n1} p_{n2}}
 {P_1^2 P_2^2 (P_1+P_2)^2}
 \;; \la{I3wn} 
\ea
analytically continuing 
$\rho_i(\omega) = \im \tilde{\mathcal{I}}_i(\omega_n\to -i [\omega + i 0^+])$; 
taking the limit $\omega\to 0$; and keeping only the terms that give
a non-vanishing contribution, some work leads to 
\ba
 \lim_{\omega\to 0^+} \frac{T\rho_1(\omega)}{\omega}
 \!\! & = & \!\!
 \frac{\nF{}(\epsilon_{p_1})\nF{}(\epsilon_{p_2})
 [1 -  \nF{}(E_{k_1})][1 - \nF{}(E_{k_2})]
 }
 {4 \epsilon_{p_1} \epsilon_{p_2}} \, 2\pi 
 \delta(\epsilon_{p_1} + \epsilon_{p_2} - E_{k_1} - E_{k_2} )
 \;, \nn \la{rho1} \\
 \lim_{\omega\to 0^+} \frac{T\rho_2(\omega)}{\omega}
 \!\! & = & \!\!
 \lim_{\omega\to 0^+} \frac{T\rho_1(\omega)}{\omega} \times
 \frac{ (\vec{p}_1+\vec{p}_2)^2 }
 {(\vec{p}_1+\vec{p}_2)^2 -(\epsilon_{p_1}+\epsilon_{p_2})^2}
 \;, \la{rho2} \hspace*{1cm} \\
 \lim_{\omega\to 0^+} \frac{T\rho_3(\omega)}{\omega}
 \!\! & = & \!\!
 \lim_{\omega\to 0^+} \frac{T\rho_1(\omega)}{\omega} \times
 \frac{ - \epsilon_{p_1} \epsilon_{p_2} }
 {(\vec{p}_1+\vec{p}_2)^2 -(\epsilon_{p_1}+\epsilon_{p_2})^2}
 \;. \la{rho3}
\ea

In the non-relativistic limit, $M \gg \pi T$, 
the subsequent spatial integrals can also be carried out. 
Indeed detailed balance,  
\ba
 && \nF{}(\epsilon_{p_1})\nF{}(\epsilon_{p_2})
 [1 -  \nF{}(E_{k_1})][1 - \nF{}(E_{k_2})] 
 \delta(\epsilon_{p_1} + \epsilon_{p_2} - E_{k_1} - E_{k_2} )
 \nn 
 & = & 
 \nF{}(E_{k_1})\nF{}(E_{k_2})
 [1 -  \nF{}(\epsilon_{p_1})][1 - \nF{}(\epsilon_{p_2})] 
 \delta(\epsilon_{p_1} + \epsilon_{p_2} - E_{k_1} - E_{k_2} )
 \;, \la{balance}
\ea
guarantees that the momenta $k_1,k_2$ are non-relativistic, 
like in \eq\nr{Boltzmann}. Momentum conservation 
requires that $\vec{p_1+p_2}$ is also non-relativistic, 
and that $\nF{}(\epsilon_{p_i})$ are exponentially small. 
So, from \eqs\nr{M2_eucl_2}--\nr{balance}, 
\ba
 \lim_{\omega\to 0^+} 
 \frac{T\rho^{(q\bar{q})}_\rmii{$\Omega$}(\omega)}{\omega}
  & \approx & 4 g^4 C_F \Nc \Nf
  \int_{\vec{p_1 p_2 k_1 k_2}}
 \hspace*{-0.5cm} \frac{\nF{}(E_{k_1})\nF{}(E_{k_2})}
 {4\epsilon_{p_1}\epsilon_{p_2}}
 \nn 
 & & \times \, 
 (2\pi)^4 \delta^{(4)} 
 (\mathcal{P}_1 + \mathcal{P}_2 - \mathcal{K}_1 - \mathcal{K}_2) \, 
 \biggl\{ \fr32 - \frac{2 \epsilon_{p_1} \epsilon_{p_2}}
 {(\epsilon_{p_1} + \epsilon_{p_2})^2}\biggr\}
 \nn 
 & \approx & 
 \frac{g^4 C_F \Nc \Nf}{M^2}
 \int_{\vec{p_1 p_2 k_1 k_2}}
 \hspace*{-0.5cm} \nF{}(E_{k_1})\nF{}(E_{k_2})
 \, 
 (2\pi)^4 \delta^{(3)} 
 (\vec{p_1} + \vec{p_2})
 \delta(2 p_1 - 2 M)
 \nn 
 & = & 
 \frac{g^4 C_F \Nc \Nf}{2\pi}
 \int_{\vec{k_1}} \nF{}(E_{k_1})
 \int_{\vec{k_2}} \nF{}(E_{k_2})
 \;. 
\ea
Dividing by $\chi^{ }_{\ff}$ from \eq\nr{chiff}, \eq\nr{Gamma_compact}
finally yields 
\be
 \Gamma^{(q\bar{q})}_\rmi{chem} \approx \frac{g^4 C_F \Nf}{8\pi M^2}
 \Bigl( \frac{MT}{2\pi} \Bigr)^{\fr32} e^{- M / T}
 \;. \la{Gamma_f} 
\ee
This agrees with \eq\nr{Gamma_heur_f}.

As far as the gluonic contributions are concerned, 
the situation is complicated by the many diagrams appearing 
in \fig\ref{fig:pert}; indeed we have checked that 
all diagram classes, with two, three
and four heavy quark propagators, need to be summed together in 
order to obtain gauge-independent results. Nevertheless, without
getting lost in excruciating detail, we can draw 
on \eqs\nr{M2_orig}, \nr{M2_eucl} to present a short but ``suggestive''
argument that things work out as before. 
For the $s$-channel process, the vacuum amplitude squared reads 
\ba
 \nsum |\mathcal{M}^{ }_1|^2_{ss} & = & 
 g^4 \tr[T_{ }^a T_{ }^b] f^{acd}f^{bcd}
 \;\;
 \mathbbm{P}_T^{\sigma\tilde{\sigma}}(\mathcal{P}_1) \,
 \mathbbm{P}_T^{\rho\tilde{\rho}}(\mathcal{P}_2)
 \nn[2mm] & & \times \, 
  \frac{ \tr[\gamma^\mu (\bsl{\mathcal{K}_1} +M )
     \gamma^\nu (\bsl{\mathcal{K}_2} -M )] }
      {(\mathcal{P}_1 + \mathcal{P}_2)^4}
 \nn & & \times \, 
 \bigl[
   \eta_{{\sigma}{\rho}}(\mathcal{P}_2 - \mathcal{P}_1)_\mu
   - \eta_{{\rho}\mu}(\mathcal{P}_1 + 2 \mathcal{P}_2)_{{\sigma}}
   + \eta_{\mu{\sigma}}(2\mathcal{P}_1 + \mathcal{P}_2)_{{\rho}}
 \bigr] 
 \nn & & \times \, 
 \bigl[
   \eta_{\tilde{\sigma}\tilde{\rho}}(\mathcal{P}_2 - \mathcal{P}_1)_\nu
    -  \eta_{\tilde{\rho}\nu}(\mathcal{P}_1 + 2 \mathcal{P}_2)_{\tilde{\sigma}}
    +  \eta_{\nu\tilde{\sigma}}(2\mathcal{P}_1 + \mathcal{P}_2)_{\tilde{\rho}}
 \bigr]
 \;. \la{M1_orig}
\ea
Here $\mathbbm{P}_T$ denotes the projector from a sum over 
the on-shell gluon polarizations, and Feynman gauge was
used for the inner gluon line. On the other hand, the gluonic 
equivalent of the process in \fig\ref{fig:eucl_details} can 
be written in Feynman gauge as 
\ba
 \delta\tilde\Omega^{(gg)}(\omega_n)
 & = & -\fr12 {g^4} \tr[T_{ }^a T_{ }^b]\, f^{acd}f^{bcd}\;  
 \nn[2mm] & & \times \, 
 \Tint{P_1 P_2 \{K_1 K_2\}}
 \hspace*{-1cm} 
 \frac{ \deltabar(\omega_n+P_1+P_2-K_1-K_2)
 \, \varepsilon_{\mu;\alpha}(P_1+P_2) \varepsilon_{\nu;\beta}(P_1+P_2)
 }{
  P_1^2 P_2^2 (K_1^2 + M^2)(K_2^2 + M^2)
 }
 \nn[2mm] & & \times \, 
 \frac{
 \tr[\gamma_\mu (i \bsl{{K}_1} +M )
      \gamma_\nu (i \bsl{{K}_2} -M )]
 }{(P_1+P_2)^4 }
  \nn & & \times \, 
 \bigl[
   \delta_{\sigma\rho}(i{P}_2 - i{P}_1)_\alpha
   - \delta_{\rho\alpha}(i{P}_1 + 2 i{P}_2)_\sigma
   + \delta_{\alpha\sigma}(2i{P}_1 + i{P}_2)_\rho
 \bigr]
 \nn & & \times \, 
 \bigl[
   \delta_{\sigma\rho}(i{P}_2 - i{P}_1)_\beta
   -  \delta_{\rho\beta}(i{P}_1 + 2 i{P}_2)_\sigma
   +  \delta_{\beta\sigma}(2i{P}_1 + i{P}_2)_\rho
 \bigr]
 \;. \la{M1_eucl}
\ea
Establishing a precise equivalence between all indices requires 
adding other gluonic contributions on both sides, but 
a comparison 
with \eqs\nr{M2_orig}, \nr{M2_eucl}, for which we 
carried out a detailed analysis, allows us to 
anticipate that things work out here as well, including 
the important factor $\fr12$ in front of the gluonic 
channels in \eq\nr{Gamma_heur_1}, clearly visible in \eq\nr{M1_eucl}.

%
\section{Discussion} 
\la{se:concl}

The question of whether or not 
heavy quarks chemically equilibrate
in heavy ion collisions is sometimes addressed by 
comparing the observed total yield with that predicted by a thermal 
distribution at the {\em final} (pionic) freeze-out temperature.   
In this paper, we have have asked whether chemical equilibrium 
could be reached earlier on, at a {\em higher} temperature. 
Since there are many heavy quarks in the initial 
state, one simply needs to get rid of some of them, to arrive 
at a thermal ensemble. The rate for this is suppressed 
by $e^{- M / T}$, which is the density of antiquarks seen by any 
given heavy quark. If this suppression can be overcome then,
for a while, heavy quarks could be part of the thermal medium, 
before re-decoupling again above  
the final pionic freeze-out, explaining why more heavy quarks
and antiquarks
are observed than is due for chemical equilibrium. 

Taking the expression from \eq\nr{Gamma_heur}; 
factorizing from it the susceptibility of \eq\nr{chiff}; normalizing
the susceptibility to its value in the massless limit, to be denoted
by $\chi^{ }_0 \equiv \Nc T^3/3$; and setting $\Nc = 3$, 
the result for the chemical equilibration rate reads
\be
 \Gamma^{ }_\rmi{chem} \;\simeq\; \frac{2\pi \alpha_s^2 T^3}{9 M^2}
 \, \biggl( \fr76 + \Nf \biggr)
 \, \frac{\chi^{ }_{\ff}}{\chi^{ }_0}
 \;. \la{Gamma_appro}
\ee
Setting furthermore $\Nf = 3$, $\alpha_s \sim 0.3$, 
$M\sim 1.5$~GeV, and estimating 
${\chi^{ }_{\ff}} / {\chi^{ }_0}$ from refs.~\cite{ding,buwu}, 
we obtain $\Gamma^{-1}_\rmi{chem} \sim 10$~fm/c at $T \sim 600$~MeV, 
and  $\Gamma^{-1}_\rmi{chem} \gsim 60$~fm/c at $T\sim 400$~MeV. 
If true, these time scales indicate that chemical equilibrium
is unlikely to be reached in current heavy ion collision
experiments, where the highest temperatures are around 
$T\sim 400$~MeV and the time scale is around $10$~fm/c. 

The estimate presented in \eq\nr{Gamma_appro} is a rough one. 
In principle, a non-perturbative value could be obtained from 
\eq\nr{Gamma_compact} through numerical lattice
Monte Carlo simulations and a subsequent analytic continuation. 
For the latter step, short-distance singularities need to be 
subtracted, as has recently been elaborated upon in connection
with other transport coefficients~\cite{Bulk_wdep,cond}. 
This task is undoubtedly a hard one: as an analysis of graph~(a)
of \fig\ref{fig:pert} shows, for $\omega \gg M$ the spectral 
function behaves as 
\be
 \rho^{ }_\rmii{$\Omega$}(\omega) \stackrel{\omega \gg M}{=} 
 \frac{g^2 C_F \Nc}{120(4\pi)^3} \,
 \bigl[ \omega^6 + \rmO(\omega^2 M^4) \bigr]
 \;, 
\ee
implying that the Euclidean correlator diverges as 
$\Omega(\tau) \sim 1 / \tau^{7}$ for $\tau \ll M^{-1}$.
To subtract this dominant and any subdominant divergences 
perturbatively, and still retain a statistically significant
signal containing the thermal physics,  
would require a very precise analysis.
(Alternatively one could start with the correlator $\Delta(\tau)$ of 
\eq\nr{Phi_tau}, although this is dominated by a constant mode, 
which poses problems for some methods of analytic continuation.)
 
Nevertheless, our non-perturbative formulation may have other
uses; for instance, it may be amenable to an order-of-magnitude 
estimate in the confined phase through chiral 
effective theories, similarly to what
has previously been achieved in the case of 
the heavy flavour kinetic equilibration 
rate~\cite{hadronic}--\cite{had3}. Possibly it could also 
be combined with  non-relativistic QCD (NRQCD) where the hard 
($ p \sim M $) momentum fields have been integrated out perturbatively.
Indeed it is possible to include the effects of $ Q \overline{Q} $
annihilation in NRQCD, through a 4-fermion interaction
in the effective Lagrangian, where the effective coupling has an
imaginary part~\cite{bodwin}. In this case one cannot consider 
$\Omega ( \tau )$ of \eq\nr{Omega_tau} 
because the chromo-electric field is hard and should have 
been integrated out; but one could
compute $ \Delta ( \tau ) $ of \eq\nr{Phi_tau} instead.

We end by remarking that whereas our non-perturbative formulation
is only valid near equilibrium, the Boltzmann description can
also be applied beyond it. Since $\Gamma^{ }_\rmi{chem}$ is 
proportional to the density of the antiquarks, 
cf.\ \eqs\nr{boltzmann}--\nr{Gamma_heur_1}, 
we may expect a correspondingly faster rate in the real world where 
the heavy antiquarks appear in overabundance. 

%
\section*{Acknowledgements}

 M.L.\ thanks Y.~Burnier for helpful discussions,
 and acknowledges partial support by the BMBF under project
 {\em Heavy Quarks as a Bridge between
      Heavy Ion Collisions and QCD}.
 We thank the Institute for Nuclear Theory at the University 
 of Washington for hospitality and the U.S.\ Department of Energy 
 for partial support during the completion of this work.


\end{document}